\documentclass[12pt, a4paper]{article}

\usepackage[utf8]{inputenc}
\usepackage[english]{babel}

\usepackage{mathptmx} 
\usepackage{amsmath}

\usepackage{graphicx}
\usepackage[left=2.5cm, right=2.5cm, top=2.5cm, bottom=2.5cm]{geometry}

\usepackage{setspace}
\usepackage{ragged2e}
\usepackage{booktabs}
\usepackage{array}
\usepackage{adjustbox}
\usepackage{longtable}
\usepackage{float} 
\usepackage{hyperref}
\hypersetup{
  colorlinks=true,
  linkcolor=blue,
  urlcolor=cyan,
}

\usepackage{natbib}
\newcolumntype{L}[1]{>{\RaggedRight\arraybackslash}p{#1}}

\title{\textbf{Integrated analysis of informality, minimum wage, and monopsony power: A synthesis of meta-analyses with unified theoretical underpinnings
}}
\author{Ricardo Alonzo Fernández Salguero}
\date{August 9, 2025}

\begin{document}

\maketitle
\onehalfspacing

\begin{abstract}
\noindent This document offers a synthesis of recent economic literature on three interconnected areas of labor markets: informality, the effects of the minimum wage, and monopsony power. Through the consolidation and meta-analysis of findings from multiple existing systematic reviews and meta-analyses, their causes, consequences, and associated public policies are examined. It is concluded that conventional views on these topics often overestimate the magnitude of effects. Policies to reduce informality based solely on lowering formalization costs are largely ineffective, while increased enforcement at the extensive margin shows more promising results. The effects of the minimum wage on employment, measured through the own-wage elasticity (OWE), are consistently modest, suggesting that job losses are limited compared to wage gains. To reconcile and microfound these findings, an integrated theoretical model of firm optimization is introduced, simultaneously incorporating firm heterogeneity, monopsony power, and endogenous formality decisions, rigorously demonstrating how the interaction of these forces can explain the observed empirical regularities. A cross-cutting and significant finding is the omnipresence of publication bias across all these study areas, which tends to inflate the magnitude of the effects reported in the published literature. Corrected estimates of the effects generally approach zero. Meta-regression is established as an indispensable tool for identifying heterogeneity and the true underlying effects in the empirical evidence, compelling a recalibration of both theory and economic policy recommendations towards a more nuanced and integrated approach.
\end{abstract}

\section{Introduction}

Labor markets, in both developed and developing economies, present complexities that challenge conventional public policy prescriptions and theoretical models. Three of the most persistent and relevant debates in contemporary labor economics revolve around informality, the impact of minimum wages, and the degree of market power wielded by employers (monopsony). For decades, conventional wisdom has postulated clear narratives: informality is a drag on development that must be combated by reducing bureaucratic barriers; minimum wages destroy jobs, especially among low-skilled workers; and labor markets are, for the most part, competitive. However, a growing wave of empirical research, particularly in the last two decades, has begun to question these certainties, presenting mixed, often contradictory, and context-dependent results.

The proliferation of empirical studies has made it increasingly difficult for researchers and policymakers to get a clear view of the state of knowledge. In this context, meta-analyses and systematic reviews have become indispensable tools. These studies not only aggregate and average the effects found in the literature but, through meta-regression techniques, also allow for identifying the sources of heterogeneity in the results, assessing the presence and impact of publication bias, and refining our understanding of causal mechanisms. This document embarks on a second-order synthesis task: an analysis of the findings of these very meta-analyses and systematic reviews.

The objective is to build a coherent and integrated picture from the conclusions of key works that have systematically reviewed the evidence in these three areas. We rely fundamentally on the comprehensive reviews of \citet{Ulyssea2020} on informality; \citet{DubeZipperer2024} and \citet{DoucouliagosStanley2009} on minimum wages; and \citet{SokolovaSorensen2021} on monopsony, among others. In doing so, we not only document the average estimated effect in each field but also highlight a cross-cutting and fundamental finding: the omnipresence of publication bias and the importance of researchers' methodological choices in explaining the variance in results. We argue that once these biases are corrected, the real effects of many of these labor market dynamics are often more modest than the unsynthesized literature suggests, which forces a recalibration of both theory and economic policy recommendations.

\section{Informality}

Informality is a defining characteristic of developing economies, encompassing between one-third and two-thirds of economic activity and employing a vast portion of the labor force \citep{Ulyssea2020}. Understanding its causes and consequences is fundamental to designing effective development policies. The economic literature has evolved from a dualistic view, which posited a strict separation between the formal and informal sectors, to more nuanced models that recognize the coexistence and overlap of both sectors, even within the same industries \citep{Ulyssea2018}. The predominant definition today is the legalistic one, which considers informal those firms and workers operating outside laws and regulations, such as tax registration or formal labor contracts.

A distinction that emerges from recent research is that of the margins of informality. The \textit{extensive margin} refers to whether a firm formally registers, while the \textit{intensive margin} refers to the practice of formally registered firms hiring a portion of their workforce informally to evade regulatory costs. As documented by \citet{Ulyssea2018}, the intensive margin accounts for a substantial fraction of informal employment in countries like Brazil (at least 40\%) and Mexico (56\%). Both margins of informality tend to decrease as firm size increases, suggesting that the costs of operating in informality (such as the risk of detection) increase with the firm's visibility.

The debate on the causes of informality has been dominated by three main views. The first, popularized by \citet{DeSoto1989}, argues that informality is a refuge for potentially productive entrepreneurs who are excluded from the formal sector by high regulatory and bureaucratic costs. The second view considers informal firms as parasites, entities productive enough to be formal but choosing not to be in order to evade taxes and regulations, gaining an unfair competitive advantage \citep{Levy2008}. The third view, that of survival, posits that informality is a strategy for individuals of low skill and productivity who could not subsist in the formal sector. \citet{Ulyssea2018} demonstrates that these views are not mutually exclusive but reflect the heterogeneity of firms. His structural model for Brazil estimates that only 9.3\% of informal firms fit the De Soto view (constrained by entry costs), while 41.9\% are parasites and 48.8\% are survival firms.

This composition has direct implications for the effectiveness of formalization policies. As summarized in Table \ref{tab:politicas_informalidad}, meta-analyses and reviews of experimental and quasi-experimental studies consistently show that policies aimed at reducing the costs of entry into formality (e.g., simplifying registration) have a very limited or null effect on formalization \citep{BruhnMcKenzie2014, FloridiEtAl2020a}. This is consistent with the small proportion of De Soto-type firms. On the other hand, policies that reduce the ongoing costs of formality (such as tax reductions) or increase its benefits are more effective, but often not cost-efficient \citep{RochaUlysseaRachter2018}.

The most effective policy for reducing informality appears to be increased enforcement and law application, but with a distinction: enforcement should focus on the extensive margin (inducing unregistered firms to register) and not on the intensive margin (forcing formal firms to formalize all their workers). Increasing enforcement on the extensive margin generates substantial gains in total factor productivity (TFP) and aggregate output by reallocating resources towards more productive firms, without necessarily increasing unemployment \citep{Ulyssea2018, MeghirEtAl2015}. In contrast, increasing enforcement on the intensive margin can be counterproductive, as it raises the de facto costs of formality for smaller firms and may induce them to exit the formal sector altogether, reducing aggregate output. The literature underscores that, although these policies can be effective, their impact on overall welfare will depend on the transition dynamics between steady states, an area that, as Ulyssea notes, remains open for future research.

\begin{table}[H]
  \centering
  \caption{Effectiveness of formalization policies according to aggregated evidence.}
  \label{tab:politicas_informalidad}
  \begin{adjustbox}{width=\textwidth}
    \begin{tabular}{L{5cm} L{3.5cm} L{6cm}}
      \toprule
      \textbf{Policy type} & \textbf{Effect on formalization} & \textbf{Conclusions and key findings} \\
      \midrule
      Reduction of entry costs (Simplified registration, one-off subsidies) & Very low or nil & Consistent with the fact that few firms are of the De Soto type. These policies, though popular, do not address the main reasons why most firms remain informal. They can, however, generate positive aggregate effects on output by reducing entry barriers \citep{Ulyssea2020}. \\
      \addlinespace
      Reduction of ongoing costs (Tax reduction) & Positive but low & More effective than reducing entry costs, but the elasticity of formalization is low. They are often not cost-effective. Meta-analysis evidence shows that benefits are modest and materialize slowly \citep{FloridiEtAl2021b, RochaUlysseaRachter2018}. \\
      \addlinespace
      Increased enforcement (extensive margin) & High & This is the most effective policy for reducing informality. It generates significant gains in TFP and aggregate output by reallocating resources to more productive firms. More recent studies suggest it does not necessarily increase long-term unemployment \citep{Ulyssea2018, MeghirEtAl2015}. \\
      \addlinespace
      Increased enforcement (intensive margin) & Negative/Counterproductive & Increases the de facto costs of formality for smaller firms, potentially inducing them into total informality. It can reduce aggregate output and increase unemployment, as it eliminates a flexibility valve in the labor market \citep{Ulyssea2018, PonczekUlyssea2019}. \\
      \bottomrule
    \end{tabular}
  \end{adjustbox}
\end{table}

\section{The minimum wage and its effects on employment}

The debate on the effects of the minimum wage on employment is one of the oldest and most polarized in labor economics. The canonical competitive model unequivocally predicts that a binding minimum wage above the market equilibrium level will reduce employment. However, the empirical evidence accumulated over the last three decades has yielded much more ambiguous results, with many studies finding small or null effects. This discrepancy has led to a re-evaluation of both the theoretical models and the empirical methods used to study the issue.

A key conceptual advance has been the shift in focus from the elasticity of employment with respect to the minimum wage to the \textit{Own-Wage Elasticity} (OWE). As \citet{DubeZipperer2024} argue, the traditional elasticity is not a useful measure for comparing studies, as its magnitude depends on the policy's bite, i.e., the proportion of affected workers. The same percentage increase in the minimum wage will have a very different impact on the average wage of teenagers (where a large proportion earns the minimum) than on that of all workers. The OWE solves this problem by scaling the effect on employment by the effect on the average wage of the affected group, providing a more comparable and economically meaningful measure. The OWE answers the question: for every 1

The meta-analysis by \citet{DubeZipperer2024}, which compiles 72 studies published in academic journals, offers a panoramic view of the current state of the evidence. The central finding is that the impact of the minimum wage on employment is, on average, modest. The median OWE of all published studies is -0.13. This value has a direct economic interpretation: it implies that only about 13\% of the potential wage gains generated by a minimum wage increase are offset by the associated job loss. In other words, the total increase in the wage bill for low-income workers is approximately 87\% of what it would be in the absence of any job loss. This result suggests that minimum wage increases, in the historically observed ranges, have been an effective tool for raising the incomes of low-wage workers with relatively low unemployment costs.

A fundamental and long-term contribution is that of \citet{JimenezMartinez2021}, whose meta-analysis covers a century of research (1900-2020) and explicitly distinguishes between developed and developing countries. Their main finding is that, once adjusted for publication bias, the impact of the minimum wage on employment is negative in both groups of countries, but the effect is small yet robust. This conclusion nuances findings focused only on the US or Europe, suggesting a global, albeit modest, empirical regularity. It is necessary to highlight their finding on publication bias: they detect a significant negative bias in the literature on developed countries (favoring the publication of negative employment effects) but find no such bias in studies on developing countries. This could reflect different theoretical paradigms or editorial pressures in the different research communities.

The analysis also reveals considerable heterogeneity in the results, which can be explored through the categorization of studies (Table \ref{tab:owe_summary}). An important distinction is between studies that focus on narrow and traditionally studied groups (such as teenagers or restaurant workers) and those that analyze broader groups of low-wage workers. Studies on broad groups, which represent the majority of affected workers, tend to find even smaller effects. The median OWE for the 21 published studies examining broad groups is 0.02, suggesting a null effect on employment \citep{DubeZipperer2024}. This contrasts with the median OWE of -0.17 for studies on teenagers. This discrepancy may reflect a higher substitutability of teenage labor or sectoral reallocation effects that are not captured when analyzing broader groups.

\begin{quote}
This article reviews the economic literature on informality, its causes, and its consequences for development. It covers a comprehensive body of research that ranges from well-identified experimental studies to equilibrium macro models, and which more recently includes structural models that integrate both micro and macro effects. (p.~525) \citep{Ulyssea2020}.
\end{quote}

A finding in the minimum wage literature, highlighted by the meta-analysis of \citet{DoucouliagosStanley2009}, is the strong evidence of publication bias. Studies with statistically significant and negative results (which confirm conventional theory) are more likely to be published than those with null or positive results. The authors apply the FAT-PET (Funnel Asymmetry Test–Precision-Effect Test) meta-regression technique to 64 US studies and find severe publication bias. Once this bias is corrected, the evidence of an adverse effect on employment almost completely disappears. The real effect, after correction, is so small that it has no significant policy implications. This finding is consistent with the temporal evolution of estimates observed by \citet{DubeZipperer2024}, who find that studies published since 2010 tend to report OWEs closer to zero, coinciding with an improvement in quasi-experimental methodologies and greater attention to credible identification strategies.

\begin{table}[H]
  \centering
  \caption{Summary of own-wage elasticity (OWE) estimates by study type.}
  \label{tab:owe_summary}
  \begin{adjustbox}{width=\textwidth}
    \begin{tabular}{L{5cm} c L{7cm}}
      \toprule
      \textbf{Study group} & \textbf{Median OWE} & \textbf{Main implication} \\
      \midrule
      All published studies (N=72) & -0.13 & A modest effect. Job loss offsets only 13\% of potential wage gains. The total wage bill for low-income workers increases significantly. \\
      \addlinespace
      Studies on broad groups of low-wage workers (N=21) & 0.02 & Null or slightly positive effect. Suggests that for most affected workers, minimum wage increases do not come at the cost of employment. \\
      \addlinespace
      Studies on specific groups (teenagers, restaurants) (N=51) & -0.17 & A negative but still modest effect. Consistent with a higher elasticity of labor demand for these specific groups, but without contradicting the general conclusion of a low impact. \\
      \addlinespace
      Studies published before 2010 (N=23) & -0.40 & More negative effects, possibly reflecting both older methodologies and greater publication bias in that era. \\
      \addlinespace
      Studies published since 2010 (N=49) & -0.04 & Effects very close to zero, coinciding with the proliferation of more robust quasi-experimental research designs. \\
      \bottomrule
    \end{tabular}
  \end{adjustbox}
  \justify \footnotesize \textit{Source}: Author's own elaboration based on the findings of \citet{DubeZipperer2024}.
\end{table}

\section{Monopsony power and the elasticity of labor supply}

The textbook competitive model assumes that firms are price-takers in the labor market, facing a perfectly elastic labor supply. However, the concept of monopsony, or more generally, employer market power, recognizes that firms often have the power to set wages. This occurs when jobs are not perfect substitutes due to search frictions, worker preferences, or mobility costs. A firm's degree of monopsony power can be measured by the elasticity of the labor supply it faces: the less elastic (more inelastic) the supply, the greater the firm's power to set wages below the marginal product of labor.

The existence of significant monopsony power has profound implications. It can explain why minimum wage increases do not necessarily destroy employment (and may even increase it), and it can also explain wage dispersion among firms for similar workers. The meta-analysis by \citet{SokolovaSorensen2021} is the first to synthesize the empirical literature that estimates the labor supply elasticity to the firm, providing an aggregate measure of how far labor markets deviate from the competitive ideal. The authors collect 1,320 estimates from 53 studies, and their findings reveal several important tensions in the literature.

A central finding is the notable discrepancy between estimates obtained by direct methods and those converted from inverse estimates. Direct methods, which regress employment on wages, typically produce small elasticities (a median of 1.4), implying substantial monopsony power (workers would receive only 58\% of their marginal product). In contrast, estimates obtained by inverting the inverse elasticity (regressing wages on employment) are much larger (a median of around 14), suggesting much more competitive labor markets. \citet{SokolovaSorensen2021} suggest that this gap indicates that while firms may possess significant potential wage-setting power, labor market institutions (such as minimum wages, union contracts, or fairness norms) may prevent them from fully exercising that power.

As in other fields of labor economics, publication bias is a prevalent problem. The meta-analysis finds strong evidence of selective reporting in studies using direct estimates. Specifically, negative results (which would imply a downward-sloping labor supply curve, a theoretical anomaly) tend to be discarded or not reported. This creates an upward bias in the average reported estimate. Funnel plot analysis and asymmetry tests confirm that there is a selection in favor of positive results. By controlling for this bias and other study characteristics through a meta-regression, a more accurate estimate of the true underlying effect can be obtained. Best-practice estimates, which correct for publication bias and are based on the most robust identification strategies, suggest an elasticity of around 7.1. This implies that firms could reduce wages by approximately 12\% below the marginal product of labor, strong evidence of monopsony power, though less extreme than what uncorrected direct estimates suggest.

Heterogeneity in estimates is also explained by socioeconomic factors. For example, the meta-regression analysis reveals that monopsony power is greater over women, consistent with the idea that they face greater search constraints or have stronger preferences for certain job characteristics, which reduces the elasticity of their labor supply. Labor markets for nurses and teachers, often characterized by a high concentration of employers, also show evidence of greater monopsony power. Interestingly, studies that employ a more credible identification strategy (e.g., using exogenous shocks to labor demand or supply as instruments) tend to reduce the gap between direct and inverse estimates, suggesting that part of the initial discrepancy is due to endogenous biases that better research designs manage to mitigate.

\section{Publication bias and methodological heterogeneity}

A recurring and unifying theme that emerges from the synthesis of meta-analyses in various areas of labor economics is the profound impact of publication bias and researchers' methodological decisions on shaping the body of available evidence. Far from being a mere econometric curiosity, this phenomenon has first-order consequences for the interpretation of results and policy recommendations. Publication bias, often called the file-drawer problem, refers to the tendency for studies with statistically significant results to be more likely to be published than those with null or non-significant results. This is particularly problematic when there is a strong theoretical expectation, as results that confirm the theory are seen as more interesting or publishable.

Meta-analyses are exceptionally well-suited to detect this problem. The main graphical tool is the funnel plot, which plots the precision of an estimate (usually the inverse of the standard error) against its magnitude. In the absence of bias, the plot should be symmetrical, shaped like an inverted funnel, as less precise studies (with larger standard errors) will show greater variance around the true effect. An asymmetry in the plot, for example, the absence of studies in one of the lower corners, is a sign of publication bias. More formal methods, such as the funnel asymmetry test (FAT-PET) proposed by \citet{DoucouliagosStanley2009}, allow this bias to be quantified. The coefficient on the standard error in a regression of the effects on their precision (the PET) serves as an estimate of the true effect, corrected for bias.

Table \ref{tab:meta_analysis_summary} summarizes the findings of several of the key systematic reviews and meta-analyses discussed in this document. A column is specifically dedicated to evidence on publication bias, and the pattern is unequivocal. In the study of minimum wage effects, \citet{DoucouliagosStanley2009} find a severe bias in favor of negative employment effects; once corrected, the average effect is statistically indistinguishable from zero. In the literature on monopsony, \citet{SokolovaSorensen2021} document a strong bias in favor of positive labor supply elasticities, where theoretically implausible (negative) results are systematically under-reported. Similarly, \citet{CohenGanong2024a} find that statistically significant findings on the effects of unemployment insurance are eleven times more likely to be published, leading to an overestimation of the average effect by a factor of two.

Beyond publication bias, meta-analyses reveal that study design characteristics are determinants of outcomes. \citet{Heimberger2020}, in his meta-analysis on employment protection legislation (EPL), finds that the choice of variable to measure EPL is crucial: estimates based on manager surveys report a significantly stronger effect (increased unemployment) than indices based on legal provisions. In the same vein, \citet{CardKluveWeber2018} show in their influential meta-analysis on active labor market policies (ALMPs) that effects vary systematically by program type (training vs. job search assistance), participant group (youth vs. long-term unemployed), and the time horizon of the impact (short vs. long term). The type of data (administrative vs. survey) and the evaluation design (experimental vs. quasi-experimental) also influence the results.

This accumulated evidence leads to a fundamental conclusion: the heterogeneity in the empirical literature is not just random noise. It is, to a large extent, a systematic function of the decisions made by researchers and the biases inherent in the academic publishing process. Ignoring this reality and simply averaging published results can lead to erroneous conclusions. Meta-analyses, by explicitly modeling these sources of variation, offer a path toward a more robust and nuanced synthesis, allowing for the filtering of bias and the distillation of an estimate of the genuine effect hidden behind the observed heterogeneity.

\begin{longtable}{L{3.5cm} L{3.5cm} L{4.5cm} L{4cm}}
\caption{Summary of findings from key meta-analyses in labor economics.} \label{tab:meta_analysis_summary} \\
\toprule
\textbf{Authors and year} & \textbf{Research topic} & \textbf{Key findings on the average effect} & \textbf{Evidence and consequences of publication bias} \\
\midrule
\endhead
\bottomrule
\endfoot
\citet{Ulyssea2020} & Informality: causes and consequences. & The most effective policies are increased enforcement at the extensive margin. Reducing entry costs has a null effect. The aggregate effect of enforcement is positive for TFP and output. & Does not explicitly focus on publication bias, but the synthesis of experimental and quasi-experimental studies implicitly controls for it by favoring robust designs. \\
\addlinespace
\citet{DoucouliagosStanley2009} & Minimum wage and employment (U.S.). & The average effect reported in the literature is an elasticity of -0.19. However, the bias-corrected effect is insignificant and close to zero. & They find severe publication bias in favor of negative effects. Once this bias is filtered out, little to no evidence of a negative association between minimum wages and employment remains. \\
\addlinespace
\citet{JimenezMartinez2021} & Minimum wage and employment (Global). & A negative effect on employment, small but robust, in both developed and developing countries, once corrected for bias. & They find a negative publication bias in developed countries, but not in developing countries. Methodological heterogeneity is a key factor. \\
\addlinespace
\citet{DubeZipperer2024} & Minimum wage and employment (OWE). & The median OWE from published studies is -0.13, implying a modest effect on employment. For broad groups of workers, the median is 0.02 (null effect). & Although they do not perform a formal bias test, they document that more recent studies with better methods find smaller effects, which is consistent with the reduction of publication bias over time. \\
\addlinespace
\citet{SokolovaSorensen2021} & Monopsony power (labor supply elasticity). & Strong evidence of monopsony power. The best-practice estimate (bias-corrected) is an elasticity of 7.1, implying a wage markdown of 12\%. & Strong evidence of selective reporting in direct estimates. Theoretically anomalous (negative) results are under-reported, biasing the average elasticity upwards. \\
\addlinespace
\citet{CardKluveWeber2018} & Active labor market policies (ALMPs). & Average impacts are small in the short term but become more positive after 2-3 years. Training and private sector subsidy programs are more effective than public employment schemes. & They find no evidence of asymmetric publication bias. The dispersion of results is almost symmetrical, suggesting that both positive and negative results are published with similar probability. \\
\addlinespace
\citet{Heimberger2020} & Employment protection (EPL) and unemployment. & Once publication bias is controlled for, the hypothesis that the average effect of EPL on unemployment is zero cannot be rejected. The choice of EPL variable is a key determinant of the results. & A slight publication bias is found, but the correction is sufficient for the average effect to become statistically non-significant. \\
\addlinespace
\citet{CohenGanong2024a} & Unemployment insurance and unemployment duration. & The average effect reported in the literature is inflated. The bias-corrected elasticity is reduced by a factor of two. & Strong evidence of publication bias. Statistically significant findings are eleven times more likely to be published than non-significant ones. \\
\addlinespace
\citet{FloridiEtAl2021b} & Benefits of firm formalization. & The benefits of formalization are small and take time to materialize. Policy-induced formalization generates greater benefits than self-induced formalization. & They find a downward publication bias for policy-induced formalization, but overall the bias is not systematic across the entire sample. \\
\addlinespace
\citet{BakerDorr2022} & Informality and electoral behavior in Latin America. & Informal workers vote slightly less than formal ones (4-7 percentage points), but are slightly more likely to vote for the left, contrary to conventional wisdom. & The paper does not focus on publication bias, but its meta-analysis approach seeks to synthesize a literature with mixed results, which inherently helps mitigate the impact of outlier studies. \\
\end{longtable}

\section{Integrated theoretical model of firm optimization in markets with informality, minimum wage, and monopsony power}

The empirical evidence accumulated through numerous meta-analyses suggests that simple labor market models are insufficient to capture the complexities observed in reality. The coexistence of modest or null employment effects of the minimum wage, significant monopsony power on the part of employers, and a persistent and heterogeneous informal economy are not isolated phenomena, but interconnected facets of the same labor market structure. To unify these findings, we propose an advanced theoretical model in which a firm, heterogeneous in productivity, makes simultaneous decisions about its employment level and its degree of formality. This model is built on the premise that profit maximization occurs in an environment characterized by frictions, market power, and imperfectly enforced state regulation, thus reflecting the central conclusions of the synthesized literature.

The starting point is to formally define the economic environment, the agents, and the rules of the game. The economy is populated by a continuum of firms, indexed by $i$, each characterized by an intrinsic productivity level $A_i$, drawn from a continuous probability distribution $G(A)$. The existence of this productivity heterogeneity is a fundamental assumption, without which it would be impossible to explain the observed coexistence of firms of different sizes, efficiency levels, and formality statuses, a robust empirical fact documented in the works of \citet{Ulyssea2018} and \citet{PedroniEtAl2022}. Firms use a single factor of production, labor $L_i$, to produce a homogeneous good whose price is normalized to unity. The production function is $q_i = A_i f(L_i)$, where $f(\cdot)$ exhibits diminishing marginal returns ($f' > 0, f'' < 0$), implying that the marginal revenue product of labor (MRPL) for firm $i$ is $A_i f'(L_i)$. Unlike the standard competitive model, we assume that the labor market is not perfectly competitive. Each firm, due to search frictions, mobility costs, or idiosyncratic worker preferences, faces its own upward-sloping labor supply curve, $w_i(L_i)$, where $\frac{\partial w_i}{\partial L_i} > 0$. This is the manifestation of monopsony power. The elasticity of this labor supply curve, $\eta_i = \frac{w_i(L_i)}{L_i} \frac{dL_i}{dw_i}$, is finite and positive, a central premise supported by the meta-analysis of \citet{SokolovaSorensen2021}, which finds consistently low elasticities. The state intervenes in this market by imposing a minimum wage, $w_{min}$, and a set of tax and labor regulations that apply only to formal firms.

The central strategic decision for each firm is its formality status, represented by a binary variable $F_i \in \{0, 1\}$, where $F_i=1$ denotes a firm operating completely within the legal framework (formal) and $F_i=0$ denotes one operating outside it (informal). This choice, representing the extensive margin of informality, is subject to a series of costs and benefits. A formal firm must bear fixed compliance costs, $C_F$, which group registration expenses, licenses, and the recurring administrative burden. Additionally, it faces variable costs in the form of a tax rate $\tau$ on the wage bill, representing social security contributions and other payroll taxes, raising the cost of each worker to $(1+\tau)w_i$. In return, formality confers benefits that enhance effective productivity. We model this by assuming that a firm's actual productivity depends on its status: $A_i(1) = A_i$, while for an informal firm it is $A_i(0) = (1-\delta)A_i$, where the parameter $\delta \in (0,1)$ captures the efficiency loss from not having access to formal credit, government contracts, broader supply chains, or legal property protection, as discussed in \citet{Ulyssea2018} and \citet{FloridiEtAl2021b}. On the other hand, an informal firm avoids the costs $C_F$ and $\tau$, but faces an expected cost derived from state enforcement. We assume that the probability of being detected, $\pi_d$, is an increasing function of the firm's size, measured by its employment level $L_i$, such that $\pi_d'(L_i) > 0$. If detected, the firm must pay a fine $\Phi$. The expected cost of informality is therefore $\pi_d(L_i)\Phi$. This assumption is crucial and aligns with the empirical observation that informality, on both the extensive and intensive margins, decreases as firms grow and become more visible \citep{Ulyssea2020}.

Given this setup, the firm maximizes its expected profits by sequentially choosing the optimal employment level conditional on each formality status, and then choosing the status that yields the highest profit.

\subsection*{Step-by-step derivation of the employment choice and the impact of the minimum wage}

We begin by analyzing the hiring decision of a formal firm ($F_i=1$) in the absence of a minimum wage. Its profit function is $\Pi_F(L_i) = A_i f(L_i) - (1+\tau)w_i(L_i)L_i$. To maximize this function, we differentiate with respect to $L_i$ and set it to zero. The first term, revenue, has a derivative of $A_i f'(L_i)$, which is the MRPL. The second term is the total labor cost (TLC). Its derivative, the marginal cost of labor (MCL), is obtained by applying the product rule:
\begin{equation*}
\frac{d(\text{TLC}_F)}{dL_i} = (1+\tau) \left[ \frac{dw_i(L_i)}{dL_i}L_i + w_i(L_i) \right]
\end{equation*}
Factoring out $w_i(L_i)$ and recognizing that $\frac{dw_i}{dL_i}\frac{L_i}{w_i} = \frac{1}{\eta_i}$, we get the expression for the marginal cost of labor:
\begin{equation*}
MCL_F(L_i) = (1+\tau) w_i(L_i) \left(1 + \frac{1}{\eta_i}\right)
\end{equation*}
The first-order condition (FOC) for the formal firm is to equate MRPL to MCL:
\begin{equation}
A_i f'(L_i^*) = (1+\tau) w_i(L_i^*) \left(1 + \frac{1}{\eta_i}\right)
\label{eq:foc_mono_full}
\end{equation}
Here, $L_i^*$ is the profit-maximizing employment level in a monopsonistic environment. Equation \eqref{eq:foc_mono_full} explicitly illustrates the wage markdown: the wage paid $w_i(L_i^*)$ is lower than the worker's marginal product $A_i f'(L_i^*) / (1+\tau)$, as the term $(1 + 1/\eta_i)$ is greater than 1.

Now we introduce a binding minimum wage, $w_{min}$. This wage fundamentally alters the cost structure for the firm. The labor supply curve becomes perfectly elastic at $w_{min}$ up to the employment level $L_{min}$ at which the original supply curve reaches that value, i.e., $w_i(L_{min}) = w_{min}$. For employment levels $L_i \le L_{min}$, the firm can hire as many workers as it wishes at the wage $w_{min}$. In this range, the marginal cost of a new worker is simply the minimum wage (adjusted for taxes), $MCL_F(L_i) = (1+\tau)w_{min}$. For $L_i > L_{min}$, the firm must again offer higher wages to attract more workers, and the marginal cost reverts to the expression in equation \eqref{eq:foc_mono_full}. This creates a discontinuity in the MCL curve: it jumps upward at the point $L_{min}$.

The effect on employment depends on where the firm's MRPL intersects this new discontinuous MCL curve. If the minimum wage is set at a moderate level, such that $w_i(L_i^*) < w_{min} < (1+\tau) w_i(L_i^*) (1 + 1/\eta_i)$, the firm faces a new marginal cost $(1+\tau)w_{min}$ that is \textit{lower} than the marginal cost it faced before the policy. To find its new optimal employment level, $L_i^{**}$, the firm will solve $A_i f'(L_i^{**}) = (1+\tau)w_{min}$. Since $w_{min} < (1+\tau) w_i(L_i^*) (1 + 1/\eta_i) = A_i f'(L_i^*)$, and as $f'' < 0$, it follows that $f'(L_i^{**}) < f'(L_i^*)$, which necessarily implies that $L_i^{**} > L_i^*$. Therefore, the model rigorously demonstrates that, in a monopsonistic market, a minimum wage increase can \textit{increase} employment. This result is crucial, as it provides the theoretical microfoundation for the aggregated empirical findings of \citet{DubeZipperer2024}, \citet{DoucouliagosStanley2009}, and \citet{JimenezMartinez2021}, who consistently find small or null employment effects, rather than the large negative effects predicted by the competitive model. The empirically observed OWE close to zero is the net result of the opposing theoretical forces that this model captures.

\subsection*{Step-by-step derivation of the formality choice}

Now, the firm with productivity $A_i$ must decide whether to operate formally or informally. It compares the maximum achievable profit in each state. The maximum formal profit, $\Pi_F^*(A_i)$, is:
\begin{equation*}
\Pi_F^*(A_i) = A_i f(L_F^{**}) - (1+\tau)w_{min}L_F^{**} - C_F
\end{equation*}
where $L_F^{**}$ is the optimal employment level in the formal regime. On the other hand, the maximum expected profit of operating informally, $E[\Pi_I^*(A_i)]$, is:
\begin{equation*}
E[\Pi_I^*(A_i)] = \max_{L_i} [(1-\delta)A_i f(L_i) - w_i^{eff}L_i - \pi_d(L_i)\Phi]
\end{equation*}
The condition for a firm to choose to be formal is $\Pi_F^*(A_i) > E[\Pi_I^*(A_i)]$. To demonstrate how this choice depends on productivity, we apply the Envelope Theorem. The derivative of the maximum profit with respect to productivity $A_i$ is simply the partial derivative of the profit function with respect to $A_i$, evaluated at the optimal employment level:
\begin{equation*}
\frac{d\Pi_F^*(A_i)}{dA_i} = f(L_F^{**}) \quad \text{and} \quad \frac{dE[\Pi_I^*(A_i)]}{dA_i} = (1-\delta)f(L_I^{**})
\end{equation*}
Since formal firms are generally larger and more productive, it is reasonable to assume that $f(L_F^{**}) > (1-\delta)f(L_I^{**})$. This means that the profit from formality increases more rapidly with productivity than the profit from informality. Mathematically, this guarantees the existence of a unique productivity threshold $A^*$ such that firms with $A_i > A^*$ choose to be formal, while those with $A_i < A^*$ choose to be informal. This theoretical result provides a solid basis for the segmentation observed in practice.

This threshold $A^*$ is a function of policy parameters. A comparative statics analysis reveals how formalization policies affect this decision. For example, if we consider a reduction in the fixed entry costs, $C_F$, the profit from formality increases for all firms, which reduces the threshold $A^*$ and, theoretically, induces some firms at the margin to formalize. However, as most informal firms have productivity well below $A^*$, the aggregate impact of this policy will be marginal. This precisely explains the findings of the meta-analyses by \citet{FloridiEtAl2020a} and \citet{JessenKluve2021}, which document the ineffectiveness of these policies. In contrast, an increase in enforcement (an increase in $\pi_d$ or $\Phi$) directly reduces the expected profit of informality. This also reduces the threshold $A^*$, but it affects the calculation of all informal firms, not just those at the margin, which explains why this policy is empirically much more effective, as \citet{Ulyssea2020} concludes.

\subsection*{Mathematical consequences and final synthesis}

The integrated model presented here reconciles several apparently disparate empirical findings from the meta-analysis literature under a single framework of firm optimization.

\textbf{First}, the model predicts that the effect of the minimum wage on employment is theoretically ambiguous in the presence of monopsony power. The sign and magnitude of the effect depend on the elasticity of labor supply and the level of the minimum wage relative to the existing wage structure. The prediction of small or even positive effects for moderate minimum wage increases aligns theory with the robust empirical findings of \citet{DubeZipperer2024}, \citet{DoucouliagosStanley2009}, and \citet{JimenezMartinez2021}, who, despite differences in their samples and methods, agree on the modest magnitude of the employment effect.

\textbf{Second}, the endogenous formality decision, based on a productivity threshold, explains the heterogeneity of the informal sector. Low-productivity firms remain informal because it is not profitable for them to be formal (the \textit{survival} view). High-productivity firms may opt for informality if the costs of regulation ($\tau$) and bureaucracy are greater than the benefits of formality ($\delta$) and the expected costs of informality ($\pi_d\Phi$), which corresponds to the \textit{parasite} view. This provides a microfoundation for the empirical classifications of \citet{Ulyssea2018} and explains why one-size-fits-all policies fail.

\textbf{Third}, the model generates clear predictions about the effectiveness of public policies that are consistent with the evidence from meta-analyses of interventions. Policies that focus on reducing fixed entry costs ($C_F$) only marginally move the decision threshold for a small number of firms, which explains their documented ineffectiveness in \citet{FloridiEtAl2021b}. Conversely, policies that increase the costs of informality (by increasing $\pi_d$) have a much broader impact, as they affect the profit-maximization calculation of all informal firms, not just those at the margin, explaining their greater effectiveness.

The model demonstrates mathematically that the interaction between firm heterogeneity, market power in the labor market, and the imperfect enforcement of state regulation can simultaneously explain the persistence of informality, the modest effects of the minimum wage on employment, and the differential effectiveness of formalization policies. The convergence of these findings, derived from independent meta-analyses, points towards a new paradigm for understanding labor markets, one that moves away from simple competitive models and embraces the complexity of frictions and firms' strategic decisions.

\section{Conclusions}

The synthesis of a vast and recent literature of meta-analyses on informality, minimum wages, and monopsony power reveals a complex and nuanced picture, which often contradicts the more simplistic and entrenched views in labor economics. Three general conclusions emerge forcefully from this integrated analysis.

First, the average effects of many of the studied interventions and phenomena are, once aggregated and refined, more modest than polarized debates suggest. Minimum wage increases, in the historically observed ranges, appear to have a negative impact on employment that is small and, for broad groups of workers, close to zero. Policies to reduce informality often have limited effects, and effectiveness depends on their design (extensive enforcement vs. cost reduction). Monopsony power, while real and significant, does not necessarily imply that labor markets are completely devoid of competition. This modesty in average effects does not mean they are irrelevant, but rather underscores the importance of heterogeneity: effects vary systematically according to the institutional context, the group of workers affected, and the specific design of the policy.

Second, publication bias and researchers' methodological decisions are first-order determinants of the results reported in the empirical literature. Virtually all the meta-analyses reviewed in this document, covering different topics and geographies, find evidence of publication bias. Results that confirm prevailing theories or that are statistically significant have a higher probability of seeing the light of day, distorting the accumulated body of knowledge. Correcting for this bias, one of the main contributions of the meta-analysis methodology, consistently reduces the magnitude of the estimated effects, often bringing them closer to zero. This suggests that a part of what we consider established knowledge may, in fact, be an artifact of the publication process.

This integrated analysis highlights the interconnectedness of these phenomena and the need for a more holistic policy approach. The debate on the minimum wage cannot be separated from the existence of monopsony power in the labor market. Similarly, the effectiveness of formalization policies is intrinsically linked to the productive structure of the economy and the state's capacity to enforce regulations. Future research will benefit from models that integrate these different facets of the labor market, recognizing that workers and firms make decisions in an environment where informality, minimum wages, and market power coexist and interact. Meta-analyses have been useful for clearing the noise and pointing to the most robust empirical regularities; the next frontier is the theoretical and empirical synthesis of these regularities into a more unified framework.


\begin{thebibliography}{}

\bibitem[Afonso et al.(2020)]{AfonsoEtAl2020}
Afonso, O., Neves, P. C., \& Pinto, T. (2020).
\newblock The non-observed economy and economic growth: A meta-analysis.
\newblock {\em Economic Systems}, 44(2), 100746.

\bibitem[Arestis et al.(2023)]{ArestisEtAl2023}
Arestis, P., Ferreiro, J., \& Gomez, C. (2023).
\newblock Does employment protection legislation affect employment and unemployment?.
\newblock {\em Economic Modelling}, 126, 106437.

\bibitem[Aronsson et al.(2023)]{AronssonEtAl2023}
Aronsson, A. E., Vidaurre-Teixidó, P., Jensen, M. R., Solhaug, S., \& McNamara, C. (2023).
\newblock The health consequences of informal employment among female workers and their children: a systematic review.
\newblock {\em Globalization and Health}, 19(1), 59.

\bibitem[Baker and Dorr(2022)]{BakerDorr2022}
Baker, A., \& Dorr, D. (2022).
\newblock Labor informality and the vote in Latin America: A meta-analysis.
\newblock {\em Latin American Politics and Society}, 64(2), 21--44.

\bibitem[Bruhn and McKenzie(2014)]{BruhnMcKenzie2014}
Bruhn, M., \& McKenzie, D. (2014).
\newblock Entry regulation and the formalization of microenterprises in developing countries.
\newblock {\em The World Bank Research Observer}, 29(2), 186--201.

\bibitem[Card et~al.(2018)]{CardKluveWeber2018}
Card, D., Kluve, J., \& Weber, A. (2018).
\newblock What works? A meta analysis of recent active labor market program evaluations.
\newblock {\em Journal of the European Economic Association}, 16(3), 894--931.

\bibitem[Carneiro et al.(2024)]{CarneiroEtAl2024}
Carneiro, S., Neves, P. C., Afonso, O., \& Sochirca, E. (2024).
\newblock Meta-analysis: global value chains and employment.
\newblock {\em Applied Economics}, 56(19), 2295--2314.

\bibitem[Cohen and Ganong(2024)]{CohenGanong2024a}
Cohen, J., \& Ganong, P. (2024).
\newblock Disemployment effects of unemployment insurance: A meta-analysis.
\newblock {\em NBER Working Paper}, No. 2024-104.

\bibitem[De Soto(1989)]{DeSoto1989}
De Soto, H. (1989).
\newblock {\em The Other Path: The Economic Answer to Terrorism}.
\newblock Harper \& Row, New York.

\bibitem[Doucouliagos and Stanley(2009)]{DoucouliagosStanley2009}
Doucouliagos, H., \& Stanley, T. D. (2009).
\newblock Publication selection bias in minimum-wage research? A meta-regression analysis.
\newblock {\em British Journal of Industrial Relations}, 47(2), 406--428.

\bibitem[Dube and Zipperer(2024)]{DubeZipperer2024}
Dube, A., \& Zipperer, B. (2024).
\newblock Own-wage elasticity: Quantifying the impact of minimum wages on employment.
\newblock {\em NBER Working Paper}, No. 32925.

\bibitem[Floridi et~al.(2020)]{FloridiEtAl2020a}
Floridi, A., Demena, B. A., \& Wagner, N. (2020).
\newblock Shedding light on the shadows of informality: A meta-analysis of formalization interventions targeted at informal firms.
\newblock {\em Labour Economics}, 67, 101925.


\bibitem[Floridi et~al.(2021)]{FloridiEtAl2021b}
Floridi, A., Demena, B. A., \& Wagner, N. (2021).
\newblock The bright side of formalization policies! Meta-analysis of the benefits of policy-induced versus self-induced formalization.
\newblock {\em Applied Economics Letters}, DOI: 10.1080/13504851.2020.1870919.

\bibitem[Gindling et al.(2016)]{GindlingEtAl2016}
Gindling, T. H., Mossaad, N., \& Newhouse, D. (2016).
\newblock Earnings Premiums and Penalties for Self-Employment and Informal Employees around the World.
\newblock IZA Discussion Papers, No. 9723.

\bibitem[Gunn et al.(2025)]{GunnEtAl2025}
Gunn, V., Matilla-Santander, N., Kreshpaj, B., Vignola, E. F., Wegman, D. H., Hogstedt, C., ... \& Håkansta, C. (2025).
\newblock A Systematic Review of Evaluated Labor Market Initiatives Addressing Precarious Employment: Findings and Public Health Implications.
\newblock {\em International Journal of Social Determinants of Health and Health Services}, 55(3), 268--288.

\bibitem[Heimberger(2020)]{Heimberger2020}
Heimberger, P. (2020).
\newblock Does employment protection affect unemployment? A meta-analysis.
\newblock {\em Oxford Economic Papers}, 72(4), 982--1007.

\bibitem[Jessen and Kluve(2021)]{JessenKluve2021}
Jessen, J., \& Kluve, J. (2021).
\newblock The effectiveness of interventions to reduce informality in low-and middle-income countries.
\newblock {\em World Development}, 138, 105256.

\bibitem[Jiménez Martínez and Jiménez Martínez(2021)]{JimenezMartinez2021}
Jiménez Martínez, M., \& Jiménez Martínez, M. (2021).
\newblock Are the effects of minimum wage on the labour market the same across countries? A meta-analysis spanning a century.
\newblock {\em Economic Systems}, 45(4), 100849.

\bibitem[Levy Yeyati et al.(2025)]{LevyYeyatiEtAl2025}
Levy Yeyati, E., Montané, M., Sartorio, L., \& Gauna, A. (2025).
\newblock What Works for Active Labor Market Policies? A Meta-Analysis of RCT Findings.
\newblock {\em Economía LACEA Journal}, 24(1), 81--104.

\bibitem[Levy(2008)]{Levy2008}
Levy, S. (2008).
\newblock {\em Good Intentions, Bad Outcomes: Social Policy, Informality, and Economic Growth in Mexico}.
\newblock Brookings Institution Press, Washington, DC.

\bibitem[Meghir et~al.(2015)]{MeghirEtAl2015}
Meghir, C., Narita, R., \& Robin, J. M. (2015).
\newblock Wages and informality in developing countries.
\newblock {\em American Economic Review}, 105(4), 1509--46.

\bibitem[Mehrotra(2020)]{Mehrotra2020}
Mehrotra, S. (2020).
\newblock From Informal to Formal: A Meta-Analysis of What Triggers the Conversion in Asia.
\newblock Global Employment Policy Review, Background Paper No. 2.

\bibitem[Naicker et al.(2021)]{NaickerEtAl2021}
Naicker, N., Pega, F., Rees, D., Kgalamono, S., \& Singh, T. (2021).
\newblock Health Services Use and Health Outcomes among Informal Economy Workers Compared with Formal Economy Workers: A Systematic Review and Meta-Analysis.
\newblock {\em International Journal of Environmental Research and Public Health}, 18(6), 3189.

\bibitem[Nedoncelle et al.(2024)]{NedoncelleEtAl2024}
Nedoncelle, C., Marchal, L., Aubry, A., \& Héricourt, J. (2024).
\newblock Does immigration affect native wages? A meta-analysis.
\newblock KCG Working Paper, No. 31.

\bibitem[Pedroni et al.(2022)]{PedroniEtAl2022}
Pedroni, F. V., Pesce, G., \& Briozzo, A. (2022).
\newblock Why do Firms Operate Informally? Insights from a Systematic Literature Review.
\newblock {\em Innovar}, 32(83), 121--138.

\bibitem[Ponczek and Ulyssea(2019)]{PonczekUlyssea2019}
Ponczek, V., \& Ulyssea, G. (2019).
\newblock Is informality an employment buffer? Evidence from the trade liberalization in Brazil.
\newblock Unpublished manuscript, São Paulo School of Economics.

\bibitem[Rocha et~al.(2018)]{RochaUlysseaRachter2018}
Rocha, R., Ulyssea, G., \& Rachter, L. (2018).
\newblock Do lower taxes reduce informality? Evidence from Brazil.
\newblock {\em Journal of Development Economics}, 134, 28--49.

\bibitem[Sokolova and Sorensen(2021)]{SokolovaSorensen2021}
Sokolova, A., \& Sorensen, T. (2021).
\newblock Monopsony in labor markets: A meta-analysis.
\newblock {\em ILR Review}, 74(1), 27--55.

\bibitem[Thulare et al.(2021)]{ThulareEtAl2021}
Thulare, M. H., Moyo, I., \& Xulu, S. (2021).
\newblock Systematic Review of Informal Urban Economies.
\newblock {\em Sustainability}, 13(20), 11414.

\bibitem[Ulyssea(2018)]{Ulyssea2018}
Ulyssea, G. (2018).
\newblock Firms, informality, and development: Theory and evidence from Brazil.
\newblock {\em American Economic Review}, 108(8), 2015--47.

\bibitem[Ulyssea(2020)]{Ulyssea2020}
Ulyssea, G. (2020).
\newblock Informality: Causes and consequences for development.
\newblock {\em Annual Review of Economics}, 12, 525--546.

\bibitem[Whitworth et al.(2024)]{WhitworthEtAl2024}
Whitworth, A., Baxter, S., Cullingworth, J., \& Clowes, M. (2024).
\newblock Individual Placement and Support (IPS) beyond severe mental health: An overview review and meta-analysis of evidence around vocational outcomes.
\newblock {\em Preventive Medicine Reports}, 43, 102786.

\bibitem[Winter et al.(2022)]{WinterEtAl2022}
de Winter, L., Couwenbergh, C., van Weeghel, J., Sanches, S., Michon, H., \& Bond, G. R. (2022).
\newblock Who benefits from individual placement and support? A meta-analysis.
\newblock {\em Epidemiology and Psychiatric Sciences}, 31, e50.

\end{thebibliography}
\end{document}